\documentclass[conference,a4paper]{IEEEtran}

\usepackage[utf8]{inputenc} 
\usepackage[T1]{fontenc}
\usepackage{url}
\usepackage{ifthen}
\usepackage{cite}
\usepackage[cmex10] {amsmath}
\usepackage{comment}
\usepackage{setspace}
\usepackage{graphicx}
\usepackage{amssymb,amsfonts}
\usepackage{bm}
\usepackage{color}

\newcommand{\matr}[1]{\mathbf{#1}} 

\begin{document}
\title{Analysis of Coded Selective-Repeat ARQ \\via Matrix Signal-Flow Graphs}


 \author{%
  \IEEEauthorblockN{Derya Malak and Muriel M\'{e}dard}
  \IEEEauthorblockA{Research Laboratory of Electronics (RLE)\\
                    Massachusetts Institute of Technology\\
                    Cambridge, MA 02139 USA\\
                    Email: \{deryam, medard\}@mit.edu}
  \and
  \IEEEauthorblockN{Edmund M. Yeh} 
  \IEEEauthorblockA{Electrical and Computer Engineering Department\\
                    Northeastern University\\ 
                    Boston, MA 02115, USA\\
                    Email: eyeh@ece.neu.edu}
}

\maketitle

\begin{abstract} 
We propose two schemes for selective-repeat ARQ protocols over packet erasure channels with unreliable feedback: (i) a hybrid ARQ protocol with soft combining at the receiver, and (ii) a coded ARQ protocol, by building on the uncoded baseline scheme for ARQ, developed by Ausavapattanakun and Nosratinia. Our method leverages discrete-time queuing and coding theory to analyze the performance of the proposed data transmission methods. We incorporate forward error-correction to reduce in-order delivery delay, and exploit a matrix signal-flow graph approach to analyze the throughput and delay of the protocols. We demonstrate and contrast the performance of the coded protocols with that of the uncoded scheme, illustrating the benefits of coded transmissions.
\end{abstract}


\section{Introduction}
\label{intro}

Automatic Repeat reQuest (ARQ) and hybrid ARQ (HARQ) methods have been used in 5G mobile networks \cite{3GPP2017}, to boost the performance of wireless technologies such as HSPA, WiMax and LTE \cite{KhosVis2017}. HARQ technique combines the important features of both forward error-correction (FEC) and ARQ error control.
A review on HARQ mechanisms that provide robustness in 4G LTE networks is given in \cite{SacBetNisApeUsh2015}. A network-coding-based HARQ algorithm for video broadcast over wireless networks is proposed in \cite{LuWuXiDu2011}. 
ARQ and HARQ protocols perform together, and provide the system with reliable packet delivery over non-deterministic channel conditions.  Here, failure in the Media Access Control 
layer HARQ operation is compensated for by the radio link control 
layer ARQ in acknowledged mode at the expense of extra 
latency for the packet \cite{3GPP2016}.

ARQ is an error-control method for data transmission that uses timeouts and acknowledgments (ACKs) to achieve reliable transmission over an erasure channel. If the sender does not receive an ACK before the timeout, it usually retransmits the packet until the sender receives an ACK or exceeds a predefined number of retransmissions. There are three basic ARQ protocols: stop-and-wait (SW) ARQ, go-back-N (GBN) ARQ, and selective-repeat (SR) ARQ \cite{AusNos2007}. These protocols use a sliding window protocol to tell the transmitter to determine which packets need to be retransmitted. 

Unreliable feedback in ARQ has been studied in \cite{KhosVis2017}, where a new method of acknowledging packet delivery for retransmission protocols is proposed. The method is based on backwards composite acknowledgment from multiple packets in a retransmission protocol, and provides the scheduler of the wireless channel with additional parameters to configure ultra-reliable communication for a user depending on channel quality. The proposed method exhibits orders of magnitude increase in reliability as compared to ARQ, at the cost of a small increase in average experienced delay.


The role of the feedback channel is to limit repetitions to only when the initial attempt fails, thereby increasing data channel efficiency. However, inevitable feedback channel impairments may cause unreliability in packet delivery. 
Attempts to increase feedback reliability, e.g., by means of repetition coding, is costly to the receiver node while erroneous feedback detection may increase packet delivery latency and diminish throughput and reliability. 
In 
LTE, blind HARQ retransmissions of a packet are proposed to avoid feedback complexity 
and increase reliability \cite{3GPP2015Mar}. This 
can severely decrease resource utilization efficiency, considering that typically a high percentage of transmissions are successfully decoded in the initial attempt in typical link adaptation configurations.

Uncoded SR ARQ protocols via signal-flow graphs have been analyzed in \cite{AusNos2007}. 
A signal-flow graph 
\cite{Howard1971} is a diagram of directed branches connecting a set of nodes. The graph represents a system of equations. The nodes are variables in the equations and the branch labels, also known as branch transmissions, represent the relationships among the variables. Scalar-flow graphs have been used to find the moment generating functions (MGFs) for the transmission and delay times in 
\cite{ChoUn1994}. Matrix signal-flow graphs (MSFGs) have been extensively used in the state-space formulation of feedback theory \cite{Chen1991}. They can be used to model channel erasures, incorporating unreliable feedback.

Different classes of codes 
have been proposed to correct errors over packet erasure channels. 
Block codes require a bit/packet stream to be partitioned into 
blocks, each block being treated independently from the rest. Block codes for error correction have been considered in \cite{KarLei2014}. 
Streaming codes, e.g. convolutional codes, have the flexibility of grouping the blocks of information in an appropriate way, and decoding the part of the sequence with fewer erasures. 
They can correct more errors than classical block codes when considering the erasure channel \cite{KarLei2014}, \cite{Lieb2017}. 
Fountain codes have efficient encoding and decoding algorithms, and are capacity-achieving. However, they are not suitable for streaming because the decoding delay is proportional to the size of the data \cite{LubMitShoSpiSte1997}.

Using FEC, in-order delivery delay over packet erasure channels can be reduced \cite{KarLei2014}, and the performance of SR ARQ protocols can be boosted. Delay bounds for convolutional codes have been provided in 
\cite{TomFitLucPedSee2014}. 
Packet dropping to reduce playback delay of streaming over an erasure channel is investigated in 
\cite{JoshiMSthesis2012}. 
Delay-optimal codes 
for burst erasure channels, and the decoding delay of codes for more general erasure models have been analyzed in \cite{Martinian2004}. 

In this paper, we use a MSFG approach to analyze the throughput and delay performance of coded SR ARQ protocols over packet erasure channels with unreliable feedback. Erasure errors can occur in both the forward and reverse channels. However, an ACK cannot be decoded as a NACK, and vice versa. We propose a SR ARQ scheme under two transmission scenarios: (i) an uncoded transmission scheme that incorporates HARQ protocol with soft combining at the receiver, and (ii) a maximum distance separable (MDS) coded ARQ scheme, by building on the uncoded baseline scheme proposed in \cite{AusNos2007}. We demonstrate the throughput and delay performance of the coded SR ARQ, and contrast it with the uncoded SR ARQ scheme. In our model, the feedback, i.e. acknowledgment (ACK) and negative acknowledgment (NACK), is cumulative, i.e. it acknowledges all the previously transmitted packets.

\section{System Model}
\label{model}  
We use a Gilbert-Elliott (GE) model 
\cite{Elliott1963}, which is a special case of hidden Markov models (HMMs), both for the forward and reverse channels, similar to the uncoded model in \cite{AusNos2007}. The status of a transmission at time $t$ is a Bernoulli random variable taking values in $\mathcal{X}=\{0,1\}$, where $0$ denotes an error-free packet, and $1$ means the packet is erroneous. This binary-state Markov process $S_t$, with probability transition matrix $\matr{P}$, has states G (good) and B (bad), i.e. $\mathcal{S}=\{G,B\}$, where $\epsilon_G$ and $\epsilon_B$ denote the probability of transmitting a bit in error in the respective states. Letting $\bm{\epsilon}=[\epsilon_G, \epsilon_B]$, the GE channel $X_t$, driven by the process $S_t$, is characterized by $\{\mathcal{S},\mathcal{X},\matr{P},\bm{\epsilon}\}$. 

For the GE channel, the state-transition matrix for both the forward and reverse channels is given by
\begin{align}
\label{TransMatrix}
\matr{P}=
\begin{bmatrix}
    1-q & q  \\
    r & 1-r
\end{bmatrix},
\end{align}
where the first and second rows correspond to states G and B. 
The stationary probabilities are
\begin{align}
\label{stationaryprobabilityGB}
\pi_G= \frac{r}{r+q},\quad \pi_B=\frac{q}{r+q}.
\end{align}
Hence, the block-error rate is $\epsilon=\pi_G \epsilon_G+\pi_B\epsilon_B$. Given $r$, $\epsilon_G$, $\epsilon_B$, and $\epsilon$, the term $q$ can be computed as $q = r \big(\frac{\epsilon_B-\epsilon_G}{\epsilon_B-\epsilon}-1\big)$. In the special case of $\epsilon_B=1$, $\epsilon_G=0$, we have $q=r\epsilon/(1-\epsilon)$. When $r=0$, the channel is memoryless, with a constant erasure rate $\epsilon=\epsilon_B$. We assume that both the forward and the reverse channels have the same parameters $r$, $\epsilon_G$, $\epsilon_B$, and $\epsilon$.

The joint probabilities of channel state and observation at time $t$, given the channel state at time $t-1$, are given as

\begin{align}
&\mathbb{P}(S_t=j, X_t=1 \vert S_{t-1}=i)\nonumber\\
&=\mathbb{P}(S_t=j \vert S_{t-1}=i)\mathbb{P}(X_t=1 \vert S_t=j)=p_{ij}\epsilon_j,\nonumber
\end{align}
which can be collected into a matrix of transition probabilities. 
The success probability matrix of the HMM is $\matr{P}_{0}^{(f)}=\matr{P}_{0}^{(r)}=\matr{P}\cdot {\rm diag}\{\matr{1}-\bm{\epsilon}\}$, 
%
%
%
and the error probability matrix of the HMM is $\matr{P}_{1}^{(f)}=\matr{P}_{1}^{(r)}=\matr{P}\cdot {\rm diag}\{\bm{\epsilon}\}$.
%
The entries in 
$\matr{P}_0$ and $\matr{P}_1$ are state-transition probabilities when viewed jointly with the conditional channel observations. 
The combined observation set is $\mathcal{X}^{(c)}=\mathcal{X}^{(f)}\times \mathcal{X}^{(r)}=\{00,01,10,11\}$, where $X_t^{(c)}=00$ means both the forward and reverse channels are good, while $X_t^{(c)}=01$ means the forward channel is good and the
reverse channel is erroneous. For $X_t^{(c)}=11$, the joint probability of the combined observation and the composite state at time $t$, given the composite state at time $t-1$, is
\begin{align}
\mathbb{P}(S_t^{(c)}&=(j,m), X_t^{(c)}=11\vert S_{t-1}^{(c)}=(i,k))\nonumber\\
&=(p_{ij}^{f}\epsilon_j^{(f)})\cdot (p_{km}^{r}\epsilon_m^{(r)}).
\end{align}
In compact notation, we have $\matr{P}_{xy}^{(c)}=\matr{P}_{x}^{(f)}	\otimes \matr{P}_{y}^{(r)}$ for $X_t^{(c)}=xy$, where $\otimes$ is the Kronecker product of matrices. 

\section{Analysis of Throughput and Delay of SR ARQ}
\label{matrixflowgraphs} 
The analysis of HMMs can be streamlined by labeling the branches of scalar-flow graphs with observation probability matrices. Flow graphs with matrix branch transmissions and vector node values are called matrix signal-flow graphs (MSFGs) \cite{AusNos2007}. The matrix gain of the graph is calculated using the basic equivalences known as parallel, series, and self-loop. Then, the desired MGF is calculated by pre- and postmultiplications of row and column vectors, respectively. 

There is a handshake mechanism between the sender and receiver that initiates a synchronous transmission. After the start of transmission (I), it takes $k-1$ time slots between the transmission of a packet and receipt of its feedback (ACK/NACK), i.e. the round-trip time (RTT) is $k$. At the transmitter, a timeout mechanism is used to prevent deadlock. When a packet is (re)transmitted, the timeout associated with this packet is set to $T$. After transmitting a new packet, the residual time for timer expiration is $d=T-k$. The timeout has to be greater than or equal to the RTT, i.e. $T\geq k$. 

The feedback includes the information about all correctly received packets\footnote{In practice, ACK/NACK do not update the transmitter about the status of any previous packets. A packet whose ACK is lost has to be retransmitted.}. The packet whose ACK is lost will be acknowledged by the subsequent ACKs/NACKs. 
If the timeout expires and no ACK is received, the packet will be retransmitted. When a packet is lost and its NACK is received, the packet will be retransmitted immediately. If the NACK is also lost, the packet will be retransmitted after the timer expires. 

\begin{figure}[t!]
\centering
\includegraphics[width=0.4\textwidth]{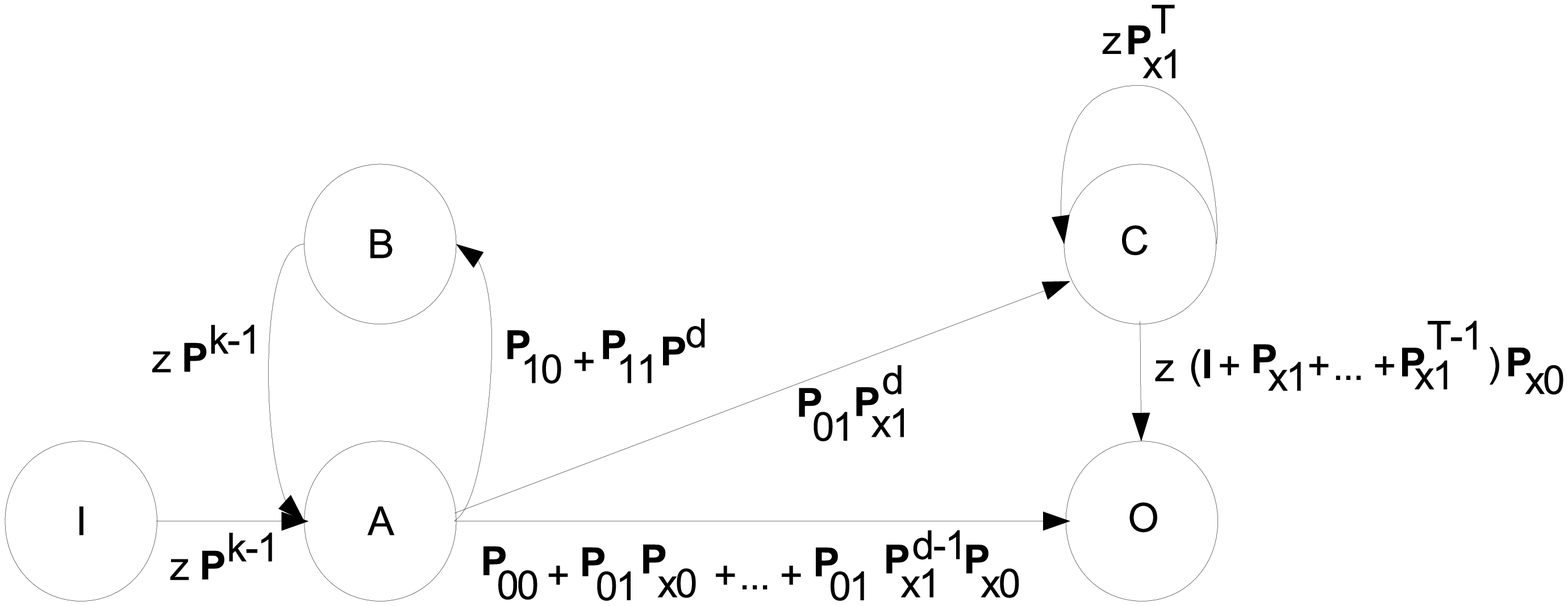}
\caption{\small{MSFG for throughput analysis of SR ARQ in unreliable feedback for the uncoded transmission scheme.}\label{NoCodingnoHARQ}}
\end{figure}

We consider a discrete-time queue method that involves a MSFG analysis for SR HARQ and coded ARQ protocols. The HMM for the uncoded SR ARQ model is illustrated in Fig. \ref{NoCodingnoHARQ}. The graph nodes correspond to the states of the transmitter. States are denoted by $I$, $A$, $B$, $C$, $O$. The input node (I) represents the start of transmission, i.e. the initial state. 
The output node (O) represents correct reception of ACK 
by the sender, i.e. the first passage time of the stochastic process. Other nodes (A, B, C) represent the hidden states. The possibilities upon the transmission of the packet are:
\begin{itemize}
\item {\bf Transition to state $A$.} After sending a new packet, the transmitter receives a feedback message $k-1$ time slots later. This state is represented by node $A$.
\item {\bf Transition to state $O$.} If the feedback is an error-free ACK, which occurs with probability $\matr{P}_{00}$, or if it is an erroneous ACK but an error-free ACK/NACK is received before timer expiration, which occurs with probability $\sum\nolimits_{k=0}^{d-1}{\matr{P}_{01}\matr{P}_{x1}^{k}\matr{P}_{x0}}$, then the system transits to state $O$ and the packet is removed from the system. 
\item {\bf Transition to state $C$.} If the feedback is an erroneous ACK and the timer expires before receiving any error-free ACKs/NACKs, the system transits to state $C$, the packet is retransmitted, modeled by the self-loop at state $C$, and the timeout is reset. The packet is acknowledged when a succeeding ACK/NACK is correctly received. 
\item {\bf Transition to state $B$.} If the feedback is an error-free NACK, which occurs with probability $\matr{P}_{10}$, or a NACK is lost and the timer expires, 
which occurs with probability $\matr{P}_{11}\matr{P}^d$, the system goes to state $B$, where the lost packet will be retransmitted. Hence, the branch gain for receiving an error-free NACK by the end of the RTT, or an erroneous NACK before the timer expires is
\begin{align}
\matr{P}_{1x}^{\rm D}=\matr{P}_{10}\matr{P}^{k-1}z^{k-1}+\matr{P}_{11}\matr{P}^{T-1}z^{T-1}.
\end{align}
\end{itemize}
The loop between states $A$ and $B$ represents retransmission of the erroneous packet until it is correctly received.

The probability vector of transmitting a new packet is $\pi_I=\pi \matr{P}_0$, where $\pi$ is the stationary vector of $\matr{P}$, 
found by solving $\pi \matr{P} = \pi$ and $\pi \matr{1} = 1$. Given the packet-error rate $\epsilon$, we have $\epsilon=\pi \matr{P}_1 \matr{1}$, and $1-\epsilon=\pi \matr{P}_0 \matr{1}$. Similar to \cite{AusNos2007}, let $\matr{P}_{0x}=\matr{P}_{00}+\matr{P}_{01}$ and $\matr{P}_{1x}=\matr{P}_{10}+\matr{P}_{11}$ be the probability matrices of success and error in the forward channel, respectively, and let $\matr{P}_{x0}=\matr{P}_{00}+\matr{P}_{10}$ and $\matr{P}_{x1}=\matr{P}_{01}+\matr{P}_{11}$ be the matrices of success and error in the reverse channel, respectively.

The transmission time $\tau$ is defined as the number of packets transmitted per successful packet, while the delay time $D$ is defined as the time from when a packet is first transmitted to when its ACK is successfully received at the sender. Both $\tau$ and $D$ are random variables with positive integer outcomes.

\paragraph{Throughput Analysis}
\label{throughputanalysis}
The MGF of the transmission time $\tau$ is calculated by left- and right-multiplying the matrix-generating function with 
$\pi_I$ and the column vector of ones:
\begin{align}
\label{generatingfunctiontransmissiontime}
\phi_{\tau}(z)=\frac{\pi_I \matr{\Phi}_{\tau}(z) \matr{1}}{\pi_I \matr{1}}
=\frac{1}{1-\epsilon} \pi \matr{P}_0 \matr{\Phi}_{\tau}(z) \matr{1}.
\end{align}
The average transmission time $\bar{\tau}$ is found by evaluating the derivative of $\phi_{\tau}(z)$ at $z=1$, as $\bar{\tau}=\phi'_{\tau}(z)\Big\vert_{z=1}$. The throughput is the reciprocal of the transmission time.

\paragraph{Delay Analysis}
\label{delayanalysis}  
The MGF of the delay $D$ is given as
\begin{align}
\label{generatingfunctiondelay}
\matr{\phi}_D(z)=\frac{\pi_{\matr{I}} \matr{\Phi}_D(z) \matr{1}}{\pi_{\matr{I}} \matr{1}}.
\end{align}
The average delay time $\bar{D}$ can be found by evaluating the derivative of $\phi_D(z)$ at $z=1$, as $\bar{D}=\phi'_{D}(z)\Big\vert_{z=1}$.

\begin{figure}[t!]
\centering
\includegraphics[width=0.4\textwidth]{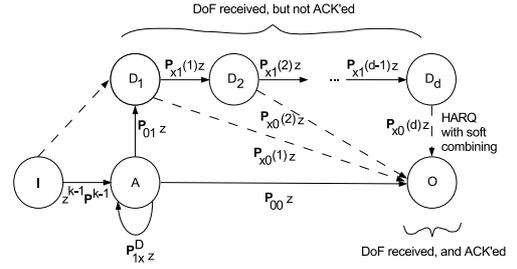}
\caption{\small{MSFG for delay analysis of SR ARQ in unreliable feedback with no coding and HARQ with soft combining at the receiver.}\label{NoCodingwithHARQ}}
\end{figure}

We skip the calculation of $\bar{\tau}$ and $\bar{D}$ for uncoded ARQ, which can be found in \cite{AusNos2007}. We next detail the HARQ protocol. 

\section{HARQ Scheme with Soft Combining}
\label{nocodingHARQ}
The HARQ protocol is an improved version of the uncoded model in \cite{AusNos2007}, in which the uncoded packets are transmitted and retransmissions are combined at the receiver. Although this scheme is suboptimal, the receiver can combine multiple transmission attempts using HARQ with soft combining, to successfully decode the transmitted packet. 
We illustrate the HMM for the HARQ transmission scheme 
in Fig. \ref{NoCodingwithHARQ}.  

\begin{figure*}[t!]
\centering
\includegraphics[width=0.73\textwidth]{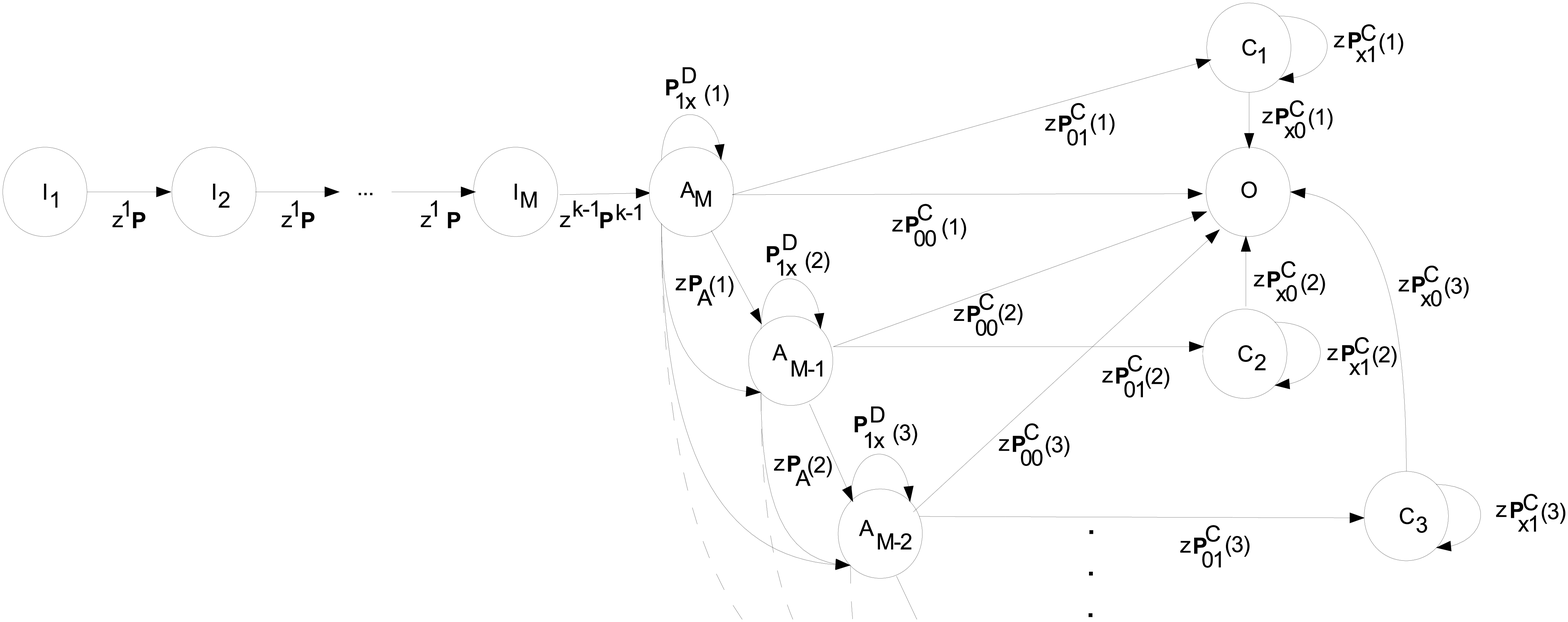}
\caption{\small{Matrix-flow graph for delay analysis of SR ARQ in unreliable feedback with coding and no soft combining at the receiver.}\label{SR_ARQ_Coding}}
\end{figure*}

The matrix-generating function of the transmission time in the case of no coding with HARQ combining is given by
\begin{align}
\label{MGFdelayNoCodingHARQ}
\matr{\Phi}_{\tau}(z)&=z\matr{P}^{k-1}(\matr{I}-z\matr{P}_{10}\matr{P}^{k-1}-z\matr{P}_{11}\matr{P}^{T-1})^{-1}\nonumber\\
&\times\left[\matr{P}_{00}+\matr{P}_{01}\sum\nolimits_{j=0}^{d-1}{\Big(\prod\nolimits_{i=0}^j{\matr{P}_{x1}(i)}\Big)}\matr{P}_{x0}(j+1)\right.\nonumber\\
&+\matr{P}_{01}\Big(\prod\nolimits_{i=0}^d{\matr{P}_{x1}(i)}\Big)
\Big(\matr{I}-z\Big(\prod\nolimits_{i=0}^{T}{\matr{P}_{x1}(i)}\Big)\Big)^{-1}\nonumber\\
&\left. z\sum\nolimits_{j=0}^{T-1}{\Big(\prod\nolimits_{i=0}^j\matr{P}_{x1}(i)\Big)}\matr{P}_{x0}(j+1)\right],
\end{align}
which is a generalization of the uncoded scheme in \cite{AusNos2007}. In the above, $\matr{P}_{x0}(i)$ and $\matr{P}_{x1}(i)$ denote the success and failure probability matrices on attempt $i\in\{1,\hdots,d-1\}$, respectively. Furthermore, $\matr{P}_{x1}(0)=\matr{I}$. 

The matrix-generating function of the delay in the case of no coding with HARQ combining is given by
\begin{eqnarray}
\label{MGFdelayNoCodingHARQ}
\matr{\Phi}_D(z)=z^{k-1}\matr{P}^{k-1}(\matr{I}-z^k \matr{P}_{10}\matr{P}^{k-1}-z^T \matr{P}_{11}\matr{P}^{T-1})^{-1}\nonumber\\
\times\left[z \matr{P}_{00}+z^2 \matr{P}_{01}   \sum\nolimits_{j=0}^{\infty}{z^j 
\prod\nolimits_{i=0}^j{\matr{P}_{x1}(i)}
}  \matr{P}_{x0}(j+1)\right].
\end{eqnarray}
%
%
Using HARQ with soft combining, we can improve the chance of successful reception. Assume that the block-error rates on a retransmission attempt $m$ for states $G$ and $B$ are
\begin{align}
\epsilon_G(m) =0,\quad
\epsilon_B(m) =1-e^{-\Gamma/(m\rho)}.\nonumber
\end{align}
The branch gain on a retransmission attempt $m$ for receiving an error-free NACK by the end of the RTT, or receiving an erroneous NACK before the timer expires equals
\begin{align}
\matr{P}_{1x}^{\rm D}(m)=\matr{P}_{10}(m)\matr{P}^{k-1}z^{k-1}+\matr{P}_{11}(m)\matr{P}^{T-1}z^{T-1},\nonumber
\end{align}
where $\matr{P}_{xy}(m)$ is the composite channel matrix for $X_m^{(c)}=xy$. Note that in the uncoded scheme, the matrices $\matr{P}_{x0}(i)$'s and $\matr{P}_{x1}(i)$'s do not change with the transmission attempt $i$. 

We next consider a coded SR ARQ protocol 
in unreliable feedback, and analyze its MSFG for throughput and delay.  
\section{Coded ARQ Scheme}
\label{coding}
In this section, we propose an MDS coded ARQ scheme. Let $\mathcal{M}=\{1,\hdots, M\}$ and $M$ be the set and the number of coded packets in the transmitted packet stream, respectively. The degrees of freedom (DoF) required at the receiver, i.e. the minimum number of independent coded packets required to reconstruct the transmitted packet stream, is $N$ packets out of $M$ packets. We do not assume in-order packet delivery. 
The transmitted stream will be successfully decoded when any subset $\mathcal{N}\subset\mathcal{M}$ of the coded transmitted packets are successfully received and acknowledged by the receiver. We assume that the feedback is cumulative.

The first feedback will be received after $k-1$ slots upon the transmission of $M$ packets. Our coded ARQ scheme is illustrated in Fig. \ref{SR_ARQ_Coding}. Node $A_M$ denotes the reception of the first feedback. The frame is retransmitted until the forward link is successful and at least one packet is transmitted. The retransmission is modeled by the self-loop at $A_M$, where
\begin{align}
\matr{P}_{1x}^{\rm D}(1)=(z\matr{P}_{10}^{\rm C}(1)+z\matr{P}_{11}^{\rm C}(1)z^{d}\matr{P}^{T-d})\matr{P}^{T-d}z^{T-d},
\end{align}
where $d=T-(k+M-1)$ is the residual time for timer expiration upon transmission, and $\matr{P}_{10}^{\rm C}(1)$ and $\matr{P}_{11}^{\rm C}(1)$ model the error-free and the erroneous NACK, respectively, upon the reception of the first feedback. If $N$ DoF's are received, the stream can be successfully decoded. If $N$ DoF's are acknowledged, 
the system transits to state $0$. Otherwise, if $N$ DoF's are received, but the feedback is an erroneous ACK, then the system transits to $C_1$, where the transmitter waits till it receives an error-free ACK/NACK. This is modeled by the self-loop at $C_1$. At node $A_M$, if the number of DoF's acknowledged by the receiver equals $m<N$, then the system transits to state 
$A_m$. 
Due to limited space, the 
expressions for transition probability matrices for $m>1$ are not given. 
 
\begin{figure*}[t!]
\centering
\includegraphics[width=0.48\textwidth]{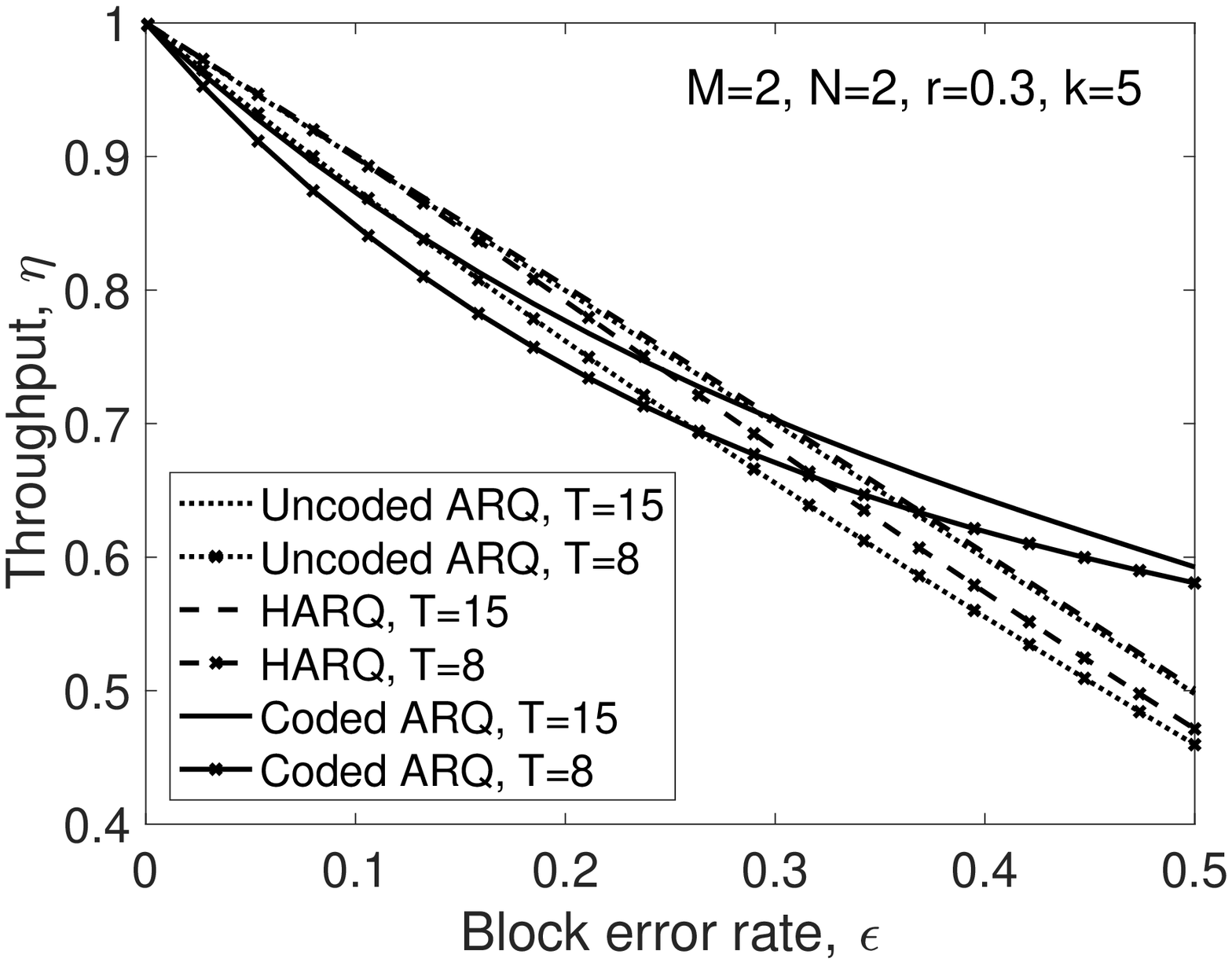}
\includegraphics[width=0.48\textwidth]{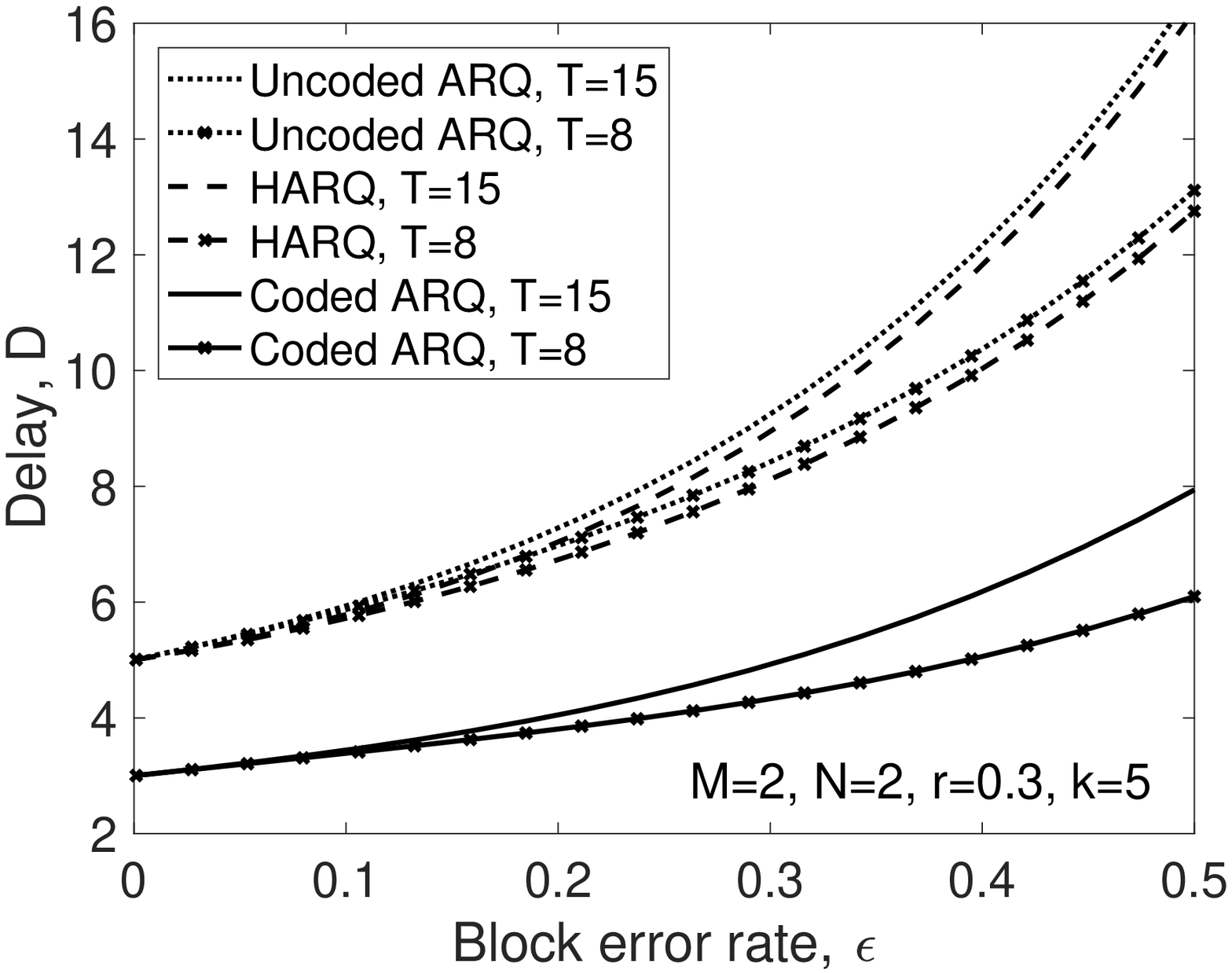}
\caption{\small{Throughput $\eta$, and average delay $\bar{D}$, versus block-error rate $\epsilon$, in Markov errors for $r=0.3$ and $k=5$.}\label{throughputdelay}}
\end{figure*}

Using Fig. \ref{SR_ARQ_Coding}, the matrix-generating function of the transmission time in the case of coded ARQ is given by
\begin{align}
\matr{\Phi}_{\tau}(z)=z\matr{P}^{k+M-2}\sum\limits_{n=1}^{\infty}{\prod\limits_{i=1}^n}{\matr{P}_A(i-1)(\matr{I}-\matr{P}_{1x}^{\mathcal{T}}(i))^{-1}A_n(z)},\nonumber
\end{align}
where $\matr{P}_A(0)=1$, and $A_n(z)$ can be computed as
\begin{align}
A_n(z)=\matr{P}_{00}^{\rm C}(n)+\matr{P}_{01}^{\rm C}(n)\left[\sum\nolimits_{i=1}^{d_n}{\matr{P}_{x1}^{\rm C}(n)^{i-1}\matr{P}_{x0}^{\rm C}(n)}\right.\nonumber\\
\left.+\matr{P}_{x1}^{\rm C}(n)^{d_n} (\matr{I}-z\matr{P}_{x1}^{\rm C}(n)^T)^{-1}z\sum\nolimits_{i=0}^{T-1}{\matr{P}_{x1}^{\rm C}(n)^i} \matr{P}_{x0}^{\rm C}(n)\right],\nonumber
\end{align}
where $d_n=d-(n-1)$. For $m\in\{1,\hdots, M\}$, 
\begin{align}
\matr{P}_{1x}^{\mathcal{T}}(m)=z(\matr{P}_{10}^{\rm C}(m)+\matr{P}_{11}^{\rm C}(m)(\matr{P}^m)^{d-m+1})\matr{P}^{T-d+m-1}.\nonumber
\end{align}
%
%
For coded ARQ, the matrix-generating function of delay is 
\begin{align}
\matr{\Phi}_D(z)=z^{k}\matr{P}^{k}\sum\limits_{n=1}^{\infty}{z^{j-1}\prod\limits_{i=1}^n \matr{P}_A(i-1)(\matr{I}-\matr{P}_{1x}^{\rm D}(i))^{-1}B_n(z)},\nonumber
\end{align}
where $B_n(z)$ can be computed using relation
\begin{align}
B_n(z)=z\matr{P}_{00}^{\rm C}(n)+z\matr{P}_{01}^{\rm C}(n)(\matr{I}-z\matr{P}_{x1}^{\rm C}(n))^{-1}z\matr{P}_{x0}^{\rm C}(n).\nonumber
\end{align}
Using 
$\matr{\Phi}_{\tau}(z)$ and $\matr{\Phi}_D(z)$, the throughput and the average delay for the coded transmission scheme can be computed. 

\section{Performance Evaluation}
\label{sims}
We evaluate the performance of the SR ARQ schemes outlined in Sects. 
\ref{nocodingHARQ}-\ref{coding} by computing the MGFs of transmission and delay times via the MSFG approach detailed in Sect. \ref{matrixflowgraphs}. The simulation parameters are selected as follows. The RTT $k=5$, and $r=0.3$ for the forward and reverse GE channels with $\epsilon_B=1$ and $\epsilon_G=0$, such that the proportion of the time spent in $G$ and $B$ can be computed using (\ref{stationaryprobabilityGB}), given the packet-error rate $\epsilon$. The performance metrics are the throughput $\eta$, which is the inverse of the average per packet transmission time, and the average per packet delay $D$ versus $\epsilon$ for varying timeout $T$. We assume the following relationship holds among the parameters: $T\geq k\geq M\geq N$. 

The throughput and delay of the different SR protocols in the Markov channel is shown in Fig. \ref{throughputdelay}. The baseline model is the uncoded ARQ scheme of \cite{AusNos2007}. For the HARQ scheme with soft combining at the receiver, $\epsilon_B(m)=1-e^{-\Gamma/(m\rho)}$, where we assume $\Gamma/\rho = 10\epsilon$, where $\rho$ is high. Hence, as $\rho$ or $m$ increases, $\epsilon_B(m)$ drops. The HARQ scheme slightly improves the delay compared to the uncoded scheme, however its throughput is similar. In the coded ARQ, more packets can be reliably transmitted even when the packet loss rate $\epsilon$ is large. As $\epsilon$ increases, throughput of coded ARQ scheme decays slower than the other schemes because coding can compensate the packet losses. Hence, less number of retransmissions is required. Furthermore, delay is significantly lower than the uncoded ARQ schemes. In all models, when the timer $T$ increases, both throughput and delay are higher. 

We leveraged discrete-time queuing and coding theory to enhance the performance of SR ARQ schemes. Contrasting the performance of HARQ with soft combining and coded ARQ with the uncoded ARQ, we demonstrated their gains in terms of throughput and delay. For the given parameter setting, the coded ARQ scheme can provide a significant reduction in delay, and a better throughput compared to the uncoded case in the high block-error rate regime. Extensions include the optimization of the erasure coded schemes with minimal encoding and decoding complexity, and the study of convolutional codes for better FEC.

\begin{spacing}{0.85}
\bibliographystyle{IEEEtran}
\bibliography{Derya}
\end{spacing}

\end{document}